\documentclass[aps,preprint,nofootinbib,superscriptaddress,prc]{revtex4}
\usepackage{amssymb}

\usepackage{epsfig}
\usepackage[bookmarksnumbered,bookmarksopen,colorlinks,citecolor=blue,linkcolor=blue]{hyperref}
\begin{document}


\title{ Shell and isospin effects in nuclear charge radii}

\author{Ning Wang}
\email{wangning@gxnu.edu.cn}
\affiliation{Department of
Physics,Guangxi Normal University, Guilin, 541004, P. R. China}

\author{Tao Li}
\affiliation{Department of Physics,Guangxi Normal University,
Guilin, 541004, P. R. China}

\fontsize{12pt}{12pt}\selectfont

\begin{abstract}
The shell effect and isospin effect in nuclear charge radii are
systematically investigated and a four-parameter formula is proposed
for the description of the root-mean-square (rms) charge radii by
combining the shell corrections and deformations of nuclei obtained
from the Weizs\"acker-Skyrme mass model. The rms deviation with
respect to the 885 measured charge radii falls to 0.022 fm. The
proposed formula is also applied for the study of the charge radii
of super-heavy nuclei and nuclear symmetry energy. The linear
relationship between the slope parameter $L$ of the nuclear symmetry
energy and the rms charge radius difference of $^{30}$S - $^{30}$Si
mirror pair is clearly observed. The estimated slope parameter is
about $L=54\pm 19$ MeV from the coefficient of the isospin term in
the proposed charge radius formula.
\end{abstract}


\maketitle

As one of the basic nuclear properties, the root-mean-square (rms)
charge radii of nuclei are of great importance for the study of
nuclear structures \cite{Ang09,Gan04} and nucleus-nucleus
interaction potentials \cite{Reis94,Zhang13}. On one hand, the rms
charge radii of nuclei can be self-consistently calculated by using
microscopic nuclear mass models such as the Skyrme
Hartree-Fock-Bogoliubov (HFB) model \cite{HFB03,HFB21}  and the
relativistic mean-field (RMF) model \cite{RMF99,Zhao10}. The HFB21
model \cite{HFB21} can reproduce the 782 measured charge radii
\cite{Ang04} with an rms deviation of 0.027 fm. On the other hand,
the rms charge radii of nuclei are also frequently described by
using mass- and isospin-dependent (or charge-dependent)
phenomenological formulas \cite{Pom94,Zhang02,Piek10,Lei09,Duflo94,Diep09}.
Although these microscopic and phenomenological models can
successfully describe the nuclear charge radii of most nuclei, the
parabolic charge radii trend in the Ca isotope chain due to the
shell closure of $N=20$ and $N=28$ cannot be reasonably well
reproduced \cite{Gan04}. The shell effect directly influences the
deformations of nuclei and thus affects the nuclear rms charge
radii. To consider the shell effect, an empirical shell correction term which is a function of the numbers
of valence nucleons was introduced in the phenomenological charge
radius formulas \cite{Ang04,Diep09}, assuming the proton magic numbers $Z_M
= 2, 6, 14, 28, 50, 82, (114)$ and neutron magic numbers $N_M=2, 8,
14, 28, 50, 82, 126, (184)$. Obviously, the fine structure of
nuclear charge radii for nuclei with semi-magic numbers such as
$Z_M=40, 64, 108$ and $N_M=56, 162$ can not be well described by
the parameterized formulas. Microscopic shell corrections and the
influence of nuclear deformations should be considered in the
formula.

In addition to the shell effect, the isospin effect also plays a
role for the nuclear charge radii. The nuclear symmetry energy, in
particular its density dependence, has received considerable
attention in recent years
\cite{Tsang09,Chen05,Shet07,Botvina02,Cent09,Stein,Stein05,Dong11}.
The nuclear symmetry energy probes the isospin part of nuclear force
and intimately relates to the structure character of neutron-rich
and neutron-deficient nuclei. The density dependence of nuclear
symmetry energy has been extensively investigated by using various
models and experimental observables, such as the microscopic
dynamics models \cite{Tsang09,Chen05}, the nuclear mass
models \cite{Wang,Wang10,Liu10,HFB21,Diep09}, the pygmy dipole
resonance \cite{Klim07,Carb10}, the neutron star observations
\cite{Stein,Wen12,Latt12}, and so on. In particular, the neutron
skin thickness of $^{208}$Pb is found to be a sensitive observable to constrain the slope parameter
$L$ of nuclear symmetry energy at the saturation density, since the
linear relationship between the slope parameter $L$ and the neutron
skin thickness $\Delta R_{\rm np}$ of $^{208}$Pb was clearly
observed \cite{Cent09,Roca}. However, it is difficult to precisely
measure the neutron radius of $^{208}$Pb in experiments, which
results in large uncertainty of the extracted slope parameter.
Comparing with the neutron radii of nuclei, the rms charge radii of
nuclei can be measured with relatively high accuracy. It would be
helpful if the slope parameter can be determined from the charge
radii of nuclei. It is known that in the absence of Coulomb
interactions between the protons, a perfectly charge-symmetric and
charge-independent nuclear force would result in the binding
energies of mirror nuclei (i.e. nuclei with the same atomic number
$A$ but with the proton number $Z$ and neutron number $N$
interchanged) being identical \cite{Lenzi,Shlomo}. At this case, the
neutron skin thickness of a neutron-rich nucleus approximately equals to the
proton radius difference (in absolute value) of this nucleus and its
mirror partner. It is therefore interesting to investigate the
correlation between the charge radius difference of mirror nuclei
and the slope parameter $L$ of the nuclear symmetry energy.

In this work, we attempt to propose a phenomenological formula for
the global description of the nuclear rms charge radii by combining
the deformations and shell corrections of nuclei obtained from the
Weizs\"acker-Skyrme mass model \cite{Wang,Wang10} which is based on
the macroscopic-microscopic method together with the Skyrme
energy-density functional and mirror constraint from the isospin
symmetry.

Based on the consideration of the nuclear saturation property, the
nuclear charge radius $R_c$ is usually described by the $A^{1/3}$
law: $R_c = r_0 A^{1/3}$, where $A$ is the
mass number. Considering the quadrupole $\beta_2$ and hexadecapole
$\beta_4$ deformations of nuclei, the rms charge radius $r_{\rm ch}$
of a nucleus can be approximately written as \cite{Zhang02}
\begin{eqnarray}
 r_{\rm ch} = \langle r^2\rangle ^{1/2}  \simeq \sqrt{\frac{3}{5}} R_c \left [1+\frac{5}{8 \pi}(\beta_2^2 +
 \beta_4^2) \right ].
\end{eqnarray}
For a better description of the charge radii of light nuclei and
nuclei far from the $\beta$-stability line, the mass- and
isospin-dependent radius coefficient $r_0$ was introduced
\cite{Pom94}, i.e. $R_c = r_0 A^{1/3}(1+\kappa/A-\alpha I)$ with the
isospin asymmetry $I=(N-Z)/A$. In addition to the mass- and
isospin-dependence of nuclear charge radii, it is found that the shell effect 
also plays a role for some nuclei \cite{Diep09}. In Fig. 1, we show the rms charge
radii of Ca isotopes and the corresponding shell corrections of
nuclei from the Weizs\"acker-Skyrme (WS*) mass model \cite{Wang10}.
One sees that the parabolic trend of the rms charge radii of Ca
isotopes between $N=20$ and 28 seems to be consistent with that of
the corresponding shell corrections, which implies that considering
the shell effect could be helpful for a better description of
nuclear charge radii.

\begin{figure} 
     \centering
        \includegraphics[width=0.75\textwidth]{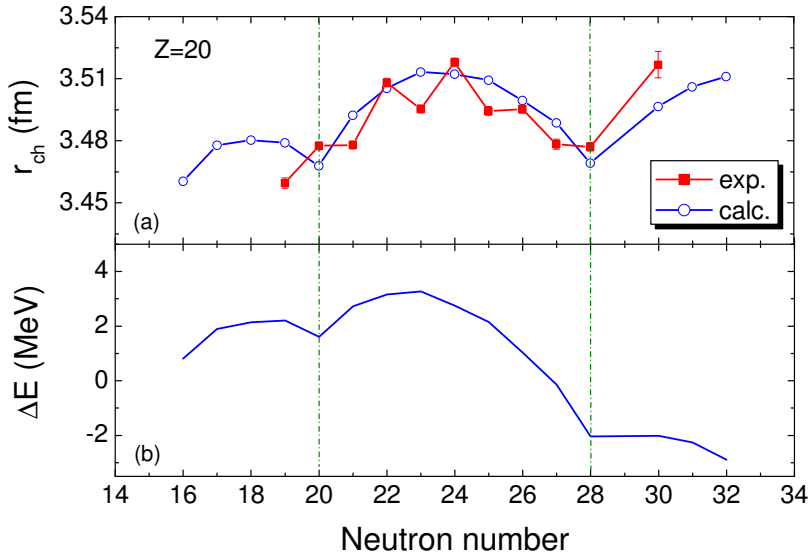}
        \caption{(Color online) (a) Nuclear rms charge radii of Ca isotopes \cite{Ang13}. (b) Shell corrections of Ca isotopes obtained from
        the  WS* mass model \cite{Wang10}. The open circles in (a)
        denote the calculated results in this work with Eqs.(1) and (2). }
    \end{figure}

Considering the influence of shell effect in nuclei, we propose a
modified four-parameter formula for the description of the nuclear
charge radius $R_c$,
\begin{eqnarray}
 R_c = r_0 A^{1/3} + r_1 A^{-2/3} + r_s I(1-I) + r_d
 \Delta E/A,
\end{eqnarray}
where $\Delta E$ denotes the shell corrections of nuclei from the
WS* mass model with which the 2149 known masses in AME2003
\cite{Audi} can be reproduced with an rms deviation of 441 keV and
the shell gaps for magic nuclei can also be well reproduced. The
$r_s$ term in Eq.(2) which is different from the isospin term in the
available phenomenological radius formulas will be discussed later.
Based on the 885 measured rms charge radii for nuclei \cite{Ang13}
with $A\geq16$ together with the deformations $\beta_2$, $\beta_4$
and shell corrections $\Delta E$ of nuclei obtained from the WS*
mass model \cite{Wang10}, and searching for the minimal rms
deviation
\begin{eqnarray}
\sigma^2=\frac{1}{m}\sum \left (r_{\rm ch}^{\rm exp} -r_{\rm
ch}^{\rm th} \right )^2
\end{eqnarray}
between the experimental data and model calculations, we obtain the
optimal values for the four parameters which are listed in Table 1.
Comparing with the rms deviation between the 885 measured radii and
the HFB21 calculations \cite{HFB21} which is 0.026 fm, the
corresponding result in this work falls to 0.022 fm.
With the microscopic shell corrections, the rms deviation of the
charge radii can be reduced by $17\%$. From Fig. 1(a), one sees that
the known rms charge radii of Ca isotopes can be reproduced
reasonably well.

\begin{table}
\caption{ Parameters of the charge radius formula $R_c$ and the rms
deviation $\sigma$ with respect to the 885 measured rms charge radii
\cite{Ang13}. The unit of $r_d$ is MeV$^{-1}$fm, and those of others
are fm.}
\begin{tabular}{ccccc}
 \hline\hline
        ~~~~~$r_0$~~~~~   & ~~~~~$r_1$~~~~~   & ~~~~~$r_s$~~~~~  & ~~~~~$r_d$~~~~~   & ~~~~~$\sigma$~~~~~ \\
\hline
 $1.2260(9)$   & $2.86(9)$ & $-1.09(3)$ & $0.99(17)$  & $0.022$ \\
  \hline\hline
\end{tabular}
\end{table}

\begin{figure} 
     \centering
        \includegraphics[width=0.75\textwidth]{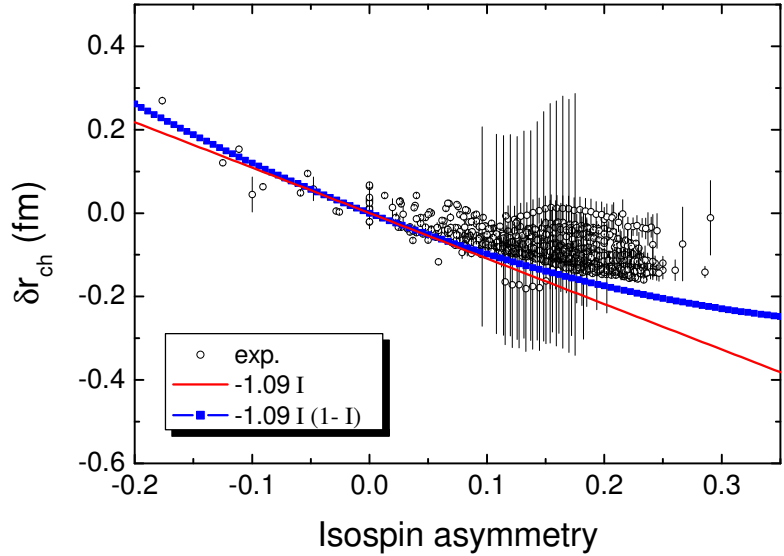}
        \caption{(Color online) Relative rms charge radii of nuclei as a function of isospin asymmetry $I$.}
    \end{figure}

For the isospin term in the charge radius formulas, the forms $I^{2}$, $I
A^{1/3}$, and $(I-I_0) A^{1/3}$  were proposed in Refs.\cite{Diep09}, \cite{Pom94} and \cite{Ang04}, respectively. Here, $I_0\simeq 0.4A/(A+200)$ denotes
the corresponding isospin asymmetry of nuclei along the
$\beta$-stability line. We note that the rms deviation can be
further reduced by about $15\%$ with a new form $I(1-I)$, comparing
with the result using the form $I A^{1/3}$. In Fig. 2, we show the
isospin dependence of the relative rms charge radii. Here the
relative rms charge radius of a nucleus is given by $\delta r_{\rm
ch}=r_{\rm ch}^{\rm exp}-\sqrt{\frac{3}{5}} \left (
r_0A^{1/3}+r_1A^{-2/3}+r_d \Delta E/A \right )$ without considering
the influence of nuclear deformations. Comparing with the linear
form, the form $I(1-I)$ gives relatively better results for the
extremely neutron-rich nuclei since the decreasing trend of the
charge radii gradually weakens with the increase of isospin
asymmetry. We also note that the value of nuclear radius constant
$r_0$ in Eq.(2) which relates to the saturation properties of
symmetric nuclear matter and neutron matter is very close to the
value (1.2257 fm) proposed in \cite{Royer09}.

\begin{figure}
     \centering
        \includegraphics[width=0.8\textwidth]{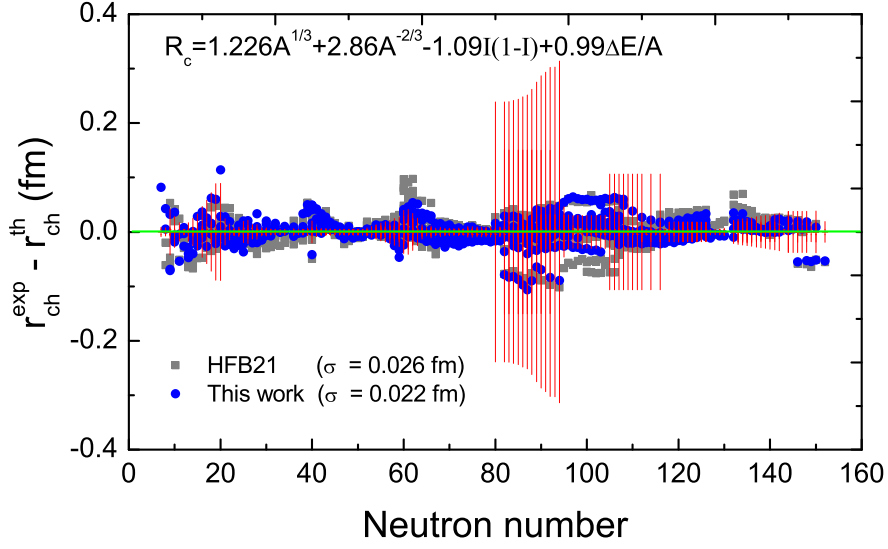}
        \caption{(Color online) Difference between the experimental data and model
calculations for the 885 rms charge radii of nuclei \cite{Ang13}.
The squares and solid circles denote the results of HFB21 and those
in this work, respectively. The error bars denote the uncertainty of
the experimental data.}
    \end{figure}

\begin{figure}
     \centering
        \includegraphics[width=0.65\textwidth]{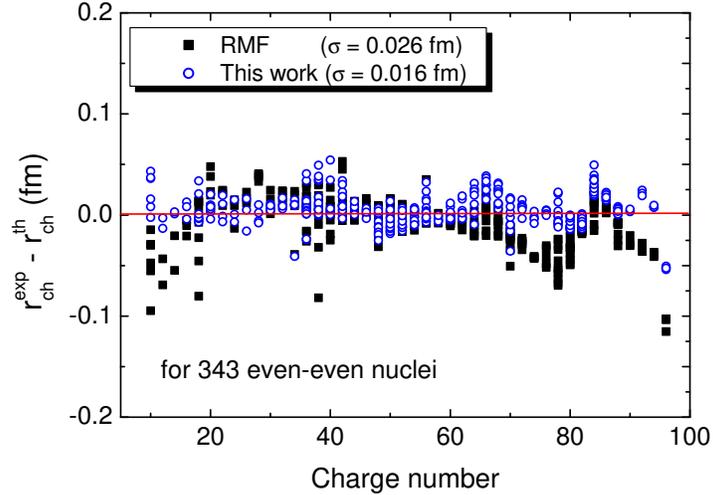}
        \caption{(Color online) Difference between the experimental data and model
calculations for the rms charge radii of 343 even-even nuclei
\cite{Ang13}. The squares and circles denote the results of RMF
model \cite{RMF99} and those in this work, respectively.  }
    \end{figure}

Based on the proposed four-parameter nuclear charge radius formula,
the known experimental data are systematically investigated. In Fig.
3, we show the difference between the experimental data and model
calculations for the 885 rms charge radii of nuclei. One sees that
the trend of the differences is similar to each other from the two
quite different models, which is due to that the obtained
deformations of nuclei with the HFB21 mass model are comparable with
the results from the WS* mass model. The calculated rms deviation
with the proposed formula is only 0.022 fm which is smaller than the
results of HFB21 by $15\%$. Here, we also present the results of the
microscopic RMF model in Fig. 4 for comparison. In Ref.
\cite{RMF99}, the rms charge radii of even-even nuclei with $Z\geq
10$ are systematically calculated by using the RMF model with the
force NL3. For the measured rms charge radii of 343 even-even
nuclei, the rms deviation with the proposed formula is 0.016 fm
which is significantly smaller than the result (0.026 fm) of the RMF
calculations. For nuclei with charge number $Z<20$ and $Z \approx 78 $, the results of
the proposed formula are better than those of the RMF calculations.
It is partly due to that the shell corrections and deformations for
nuclei with new magic numbers such as $N=14$, 16 and for nuclei with
sub-shell closure are reasonably well described by the WS* mass
model.

\begin{figure}
     \centering
        \includegraphics[width=0.9\textwidth]{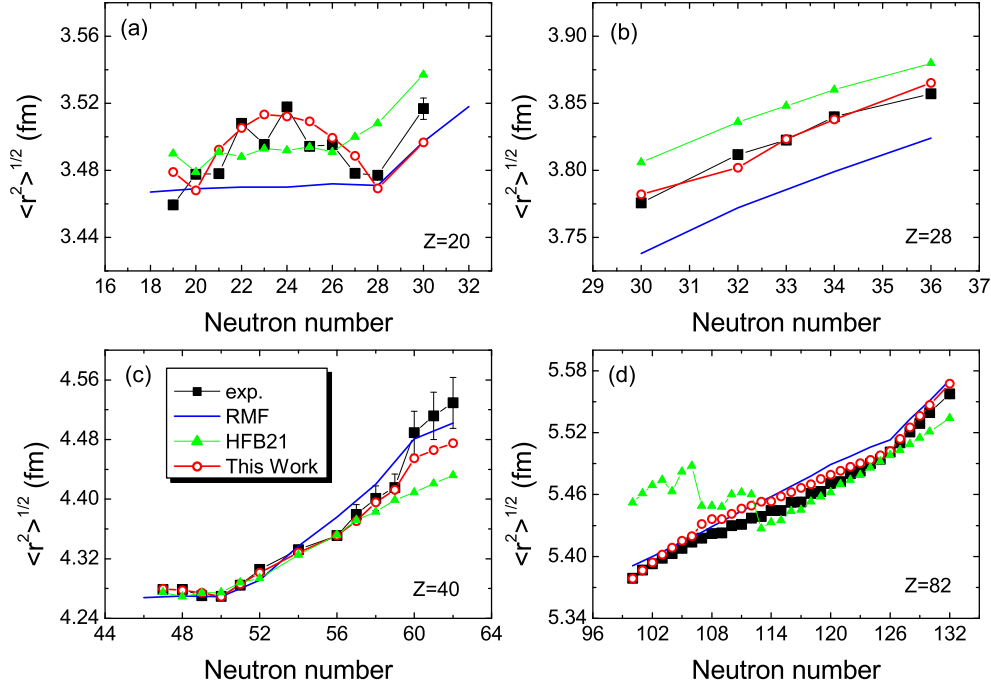}
        \caption{(Color online) Comparisons of the rms charge radii for Ca, Ni, Zr and Pb isotopes.
        The triangles, solid curves and solid squares denote the results of HFB21, RMF and experimental data, respectively.
        The open circles denote the results in this work according to Eqs.(1) and (2).  }
        \label{fig3}
    \end{figure}

In Fig. 5, we show the comparisons of the rms charge radii for Ca,
Ni, Zr and Pb isotopes from three models. For the Ca isotopes,
neither the HFB21 model nor the RMF model reproduce the trend of the
experimental data. For the Ni isotopes, the experimental data are
systematically over-predicted by the HFB21 model and under-predicted
by the RMF model. For the doubly-magic nucleus $^{56}$Ni, the
calculated quadrupole deformation of nucleus $\beta_2=0.16$ with the
HFB21 model. For other Ni isotopes, the obtained $\beta_2$ values
from the HFB21 calculations are significantly larger than the
results of WS* which results in the over-predicted results for the
Ni isotopes. According to the RMF calculations, we note that the
binding energies of $^{58,60,62,64}$Ni are systematically
under-predicted by about 4 MeV which might affect the reliable
description of the rms charge radii.  For the Zr isotopes, one sees
that the rms charge radii for nuclei with $N=50$, 56 and 58 can be
remarkably well reproduced with the proposed formula since the shell
corrections and deformations of these nuclei (with shell or
sub-shell closure) are reasonably well described by the WS* mass
model. For the neutron-deficient Pb isotopes, the rms charge radii
are significantly over-predicted by the HFB21 calculations and the kink at $N=126$ can not be reproduced. The
global trend of the rms charge radii for the nuclei in Fig. 5 especially the kinks at the magic numbers
can be well described by using the proposed nuclear charge radius formula.

\begin{figure}
     \centering
        \includegraphics[width=0.75\textwidth]{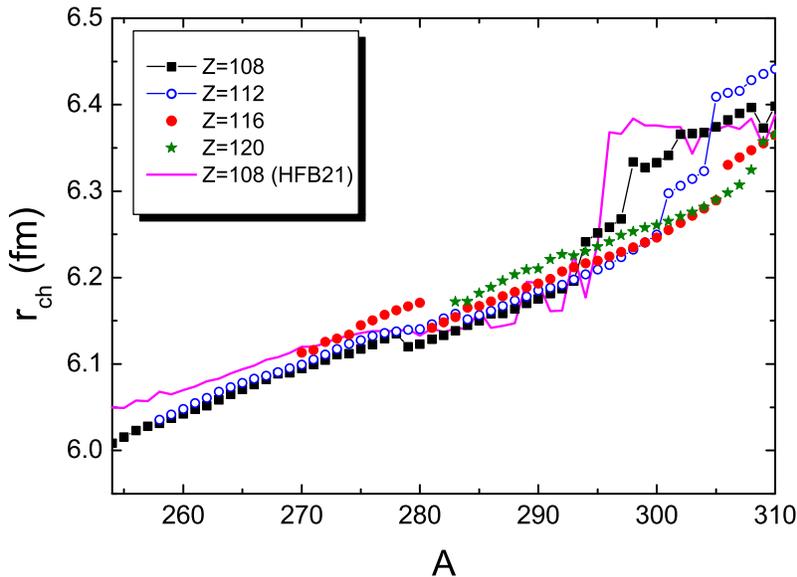}
\caption{(Color online) Predicted rms charge radii with the proposed
formula for some super-heavy nuclei. The solid curve denote the
results of HFB21 for Hs (Z=108) isotopes.}
    \end{figure}

In Fig. 6, we show the predicted rms charge radii with the proposed
formula for some super-heavy nuclei. The solid curve denote the
results of HFB21 for Hs ($Z=108$) isotopes which are comparable with
the predictions of this work (the deviations are smaller than 0.05
fm in general). For the super-heavy nuclei $^{286}$114 and
$^{290}$116, the extracted rms charge radii from the experimental
$\alpha$-decay data are $r_{\rm ch}=6.24 \pm 0.14$ and  $6.13 \pm
0.16$ fm \cite{Ni13}, respectively. The predicted results in this
work for these two nuclei are 6.17 and 6.19 fm, respectively, which
are comparable with the extracted results in \cite{Ni13}.

\begin{figure}
     \centering
        \includegraphics[width=0.75\textwidth]{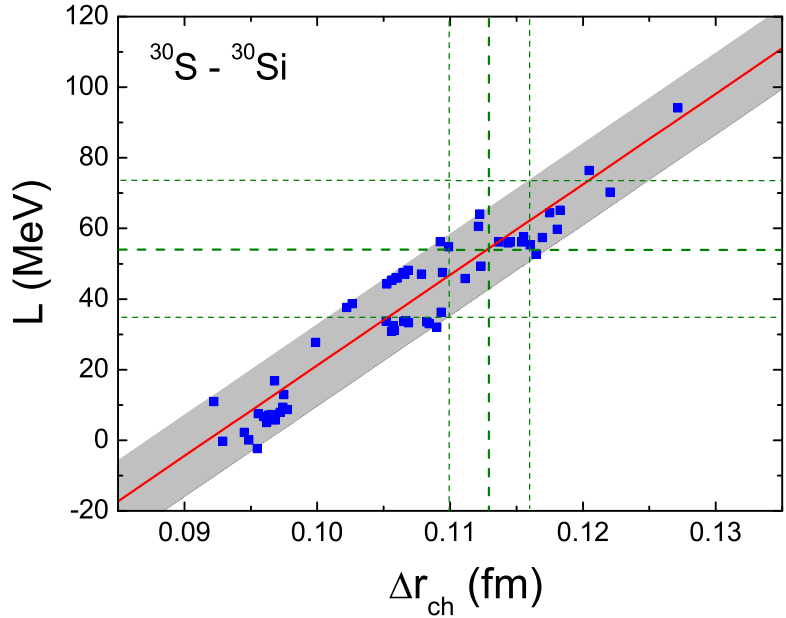}
\caption{(Color online) Slope parameter of nuclear symmetry energy
as a function of rms charge radius difference between $^{30}$S and
$^{30}$Si. The squares denote the calculated results by using the
Skyrme Hartree-Fock model with 62 different Skyrme forces. The solid
line with shadows is a linear fit to the squares. The dashed
horizontal lines denote the estimated slope parameter according to
the $r_s$ value of the proposed charge radius formula.}
     \end{figure}

In this work, we simultaneously investigate the correlation between
the nuclear symmetry energy and the isospin term of the charge
radius formula. We study the difference of the rms charge radii
between mirror nuclei, such as $^{30}$S - $^{30}$Si pair. Because of
the influence of new magic numbers $N=14$, 16, the calculated
deformations of these two nuclei are very small with some different
mass models such as the Weizs\"acker-Skyrme mass model, the finite
range droplet model \cite{Moll95}, and the HFB calculations
\cite{HFB03} adopting three widely used Skyrme forces SLy4, SkP and
SkM*. We systematically calculate the difference of the rms charge
radii $\Delta r_{\rm ch}$ between $^{30}$S and $^{30}$Si by using
the Skyrme Hartree-Fock model with 62 different Skyrme forces in
which the corresponding incompressibility coefficient for symmetry
nuclear matter $K_\infty=210 - 280$ MeV and the saturation density
$\rho_0=0.15 - 0.17$ fm$^{-3}$. From Fig. 7, one can clearly see the
linear relationship between the slope parameter $L$ and the
difference of the rms charge radii $\Delta r_{\rm ch}$. The Pearson's (linear) correlation coefficient $r$ of $L$ with $\Delta r_{\rm ch}$ 
for the 62 Skyrme forces is 0.95, which is comparable with the corresponding 
value (0.97) of $L$ with the neutron skin thickness of $^{208}$Pb for the same forces. The values of $\Delta r_{\rm ch}$ for measured mirror pairs such as $^{34}$S - $^{34}$Ar and $^{18}$O - $^{18}$Ne are also investigated with the deformed configurational Skyrme Hartree-Fock calculations \cite{Sto05}. We note that the obtained linear correlation coefficient $r$ for these two pairs are 0.58 and 0.69, respectively, which are significantly smaller than the value for the mirror pair $^{30}$S - $^{30}$Si. It indicates that the linear relationship between $L$ and $\Delta r_{\rm ch}$ for the two pairs $^{34}$S - $^{34}$Ar and $^{18}$O - $^{18}$Ne are not as good as that for the mirror pair $^{30}$S - $^{30}$Si due to the influence of the new magic numbers $N=14$, 16. Considering the fact that the experimental uncertainty
for the rms charge radius measurement is much smaller than that for
the rms neutron radius, precise measurements of the rms charge radii
for the pair of mirror nuclei $^{30}$S - $^{30}$Si, especially the
unmeasured $^{30}$S, could be very helpful for the extraction of the
slope parameter.

Due to the perfectly charge-symmetric and charge-independent nuclear
force, the deformations and shell correction of a nucleus
approximately equal to those of its mirror nucleus
\cite{Set12,Wang10}. Based on the proposed nuclear charge radius
formula, one obtains $\Delta r_{\rm ch}\approx \sqrt{\frac{3}{5}}\,
2  r_s I $ neglecting the influence of nuclear deformations. For the
$^{30}$S - $^{30}$Si mirror pair, the estimated value of $\Delta
r_{\rm ch}$ is about $0.113\pm 0.003$ fm and the corresponding slope parameter is
about $L= 54 \pm 19$ MeV which is consistent with recently extracted
results from the Fermi-energy difference in nuclei \cite{Wang13} and
from modeling X-ray bursts and quiescent low-mass X-ray binaries
\cite{Ste10,Heb13}. In addition, the extracted slope parameter $L=
52.5 \pm 20$ MeV from the Skyrme Hartree-Fock calculations together
with the neutron skin thickness of Sn isotopes \cite{Chen11} and $L=
52.7 \pm 22.5$ MeV from the global nucleon optical potentials
\cite{Xu10} are in good agreement with the estimated result in this work.

In summary, by combining the Weizs\"acker-Skyrme mass model, we
propose a four-parameter nuclear charge radius formula in which the
microscopic shell correction and nonlinear isospin terms are
introduced. The 885 measured rms charge radii of
nuclei are reproduced with an rms deviation of 0.022 fm. For the
measured even-even nuclei, the rms deviation is only 0.016 fm. The
parabolic charge radii trend in Ca chain due to the shell effect and
the trend of Ni, Zr and Pb isotopes are reasonably well described
with the formula. Through a study of the difference of the rms
charge radii $\Delta r_{\rm ch}$ between mirror nuclei
by using the Skyrme Hartree-Fock model with
62 different Skyrme forces, the linear relationship between the
slope parameter $L$ and $\Delta r_{\rm ch}$ for the mirror pair $^{30}$S -
$^{30}$Si is clearly observed, which would be helpful
for the extraction of the slope parameter of the nuclear symmetry
energy. The estimated slope parameter from the coefficient $r_s$ of
the isospin term in the proposed formula is about $L=54\pm 19$ MeV.

\begin{acknowledgments}
This work was supported by the National Natural Science Foundation
of China No. 11275052.
\end{acknowledgments}

\end{document}